\documentstyle[12pt]{article}
\def\sect#1{\section*{#1}\addtocounter{section}{1}\setcounter{equation}{0}}

\begin{document}

\pagestyle{empty}
\begin{flushright} KAIST/THP-96/702 \end{flushright}
\vspace{0.1cm}
\begin{center}
{\large {\bf Square Root Singularity in Boundary Reflection Matrix} }
\end{center}
\vspace{0.1cm}
\begin{center}
{\bf J. D. Kim\footnote{jdkim@sorak.kaist.ac.kr}} and {\bf I. G. Koh} \\
\vspace{0.5cm}
{\it Department of Physics    \\
Korea Advanced Institute of Science and Technology        \\
Taejon, 305-701 Korea}  \\ 
\end{center}
\vspace{2.5cm}

\begin{abstract}
Two-particle scattering amplitudes for integrable relativistic quantum
field theory in 1+1 dimensions can normally have at most singularities
of poles and zeros along the imaginary axis in the complex rapidity plane.
It has been supposed that single particle amplitudes of the exact
boundary reflection matrix exhibit the same structure.
In this paper, single particle amplitudes of the exact boundary
reflection matrix corresponding to the Neumann boundary condition
for affine Toda field theory associated with twisted affine algebras
$a_{2n}^{(2)}$ are conjectured, based on one-loop result,
as having a new kind of square root singularity.
\end{abstract}

\newpage
\pagestyle{plain}

\sect{1. Introduction}
Two-particle scattering amplitudes
for massive integrable relativistic quantum field theory
in 1+1 dimensions can have ordinary square root threshold singularity
in terms of Mandelstam variables\cite{ELOP}.
However, a reparametrisation of the energy-momentum of on-shell
states in terms of rapidity parameter unfolds the square root threshold
singularity so that the scattering amplitudes can have
at most zeros and poles along the imaginary axis in the complex rapidity
plane, provided they do not exhibit any other sort of branch cuts
as is usually supposed\cite{ZZ}.
In this case, odd-order poles in the physical strip are to be
interpreted as a signal of the existence of virtual bound states, which
should be among the initial spectrum of asymptotic states
by bootstrap principle.
Various aspects of the singularity structure of
the proposed exact $S$-matrices for affine Toda field theory(ATFT)
have been extensively studied in terms of the so-called Landau singularity
of Feynman diagrams\cite{BCDS,DGZ,BS,BCKKS}.

By the way, the models defined on a space with a boundary,
let say on a half line, naturally lead one to consider
the boundary reflection matrix in order to describe the reflection
process of particles against the boundary.
Indeed, since the boundary bootstrap equation\cite{FK} and the boundary
crossing-unitarity relation\cite{GZ} were introduced, a variety of solutions
of the boundary Yang-Baxter equation for the boundary reflection matrix
which was first introduced in \cite{Che} has been
constructed\cite{FK,GZ,Sas,AK}. In this algebraic approach,
$S$-matrices are used as a part of input data and proper interpretations
in the framework of Lagrangian quantum field theory was not 
given\footnote{There are some works
which aim to relate physical parameters in the boundary potential to
formal parameters arising from solutions of the algebraic
equations; for the sine-Gordon theory at a generic point in semi-classical
analysis\cite{SSW} and at the free fermion point\cite{AKL} where one may
use the method\cite{GZ} of mode expansion for the field as an operator.}.

Classical boundary reflection matrices corresponding to the various choices
of the integrable boundary condition\footnote{For studies on the possible
integrable boundary potentials,
see \cite{GZ,CDRS,CDR,Mac,BCDR,PZ,MS,IOZ,PRZ}.} have been constructed
by linearising the equation of motion around a background solution
in \cite{CDRS,CDR}, where some conjectures on the corresponding
exact boundary reflection matrices have been also made.
A study on the boundary reflection matrix in quantum field theory
has been initiated in the framework of the Feynman's perturbation theory
in \cite{Kim1} and single particle reflection amplitudes
for ATFT with the Neumann boundary condition were constructed
in \cite{Kim2,Kim3}.
Quite recently, a geometric expression of the boundary reflection matrices
in terms of root systems
for simply-laced ATFT was otained in \cite{KY}.

Single particle amplitudes of the exact boundary reflection matrix
have been usually supposed to exhibit the same analytic structure
as two-particle scattering amplitudes in the absence of a new idea.
In the mean time, there appeared a first example of single particle
amplitudes of the exact boundary reflection matrix having a new kind of
square root singularity in the case of $a_2^{(2)}$ theory\cite{Kim1}.
In the present paper, for ATFT associated with twisted affine algebras
$a_{2n}^{(2)}$ for any $n$, single particle amplitudes of the exact
boundary reflection matrix corresponding to the Neumann boundary condition
are constructed and checked, based on one-loop result,
as having square root singularities.

The plan of this paper is as follows.
Section 2 contains partial results of the single particle amplitudes
at one-loop order for $a_{2n}^{(2)}$ ATFT defined
on a half line with the Neumann boundary condition.
In section 3, a complete set of the exact boundary reflection matrix
having square root singularities is presented based on the
reduction idea\cite{OT,CM,Sas2} and checked against the one-loop result.
Finally, conclusions are made in section 4.

\sect{2. Perturbative Boundary Reflection Matrix}
The action for ATFT\cite{BCDS2} associated with a Lie algebra $g$ with
the rank $r$ defined on a half line ($ -\infty < x \leq 0$) is given by
\begin{equation}
S(\Phi) = \int_{-\infty}^{0} dx \int_{-\infty}^{\infty} dt
\left ( \frac{1}{2}\partial_{\mu}\phi^{a}\partial^{\mu}\phi^{a}
-\frac{m^{2}}{\beta^{2}}\sum_{i=0}^{r}n_{i}e^{\beta \alpha_{i} \cdot \Phi}
\right )
\end{equation}
where $ \alpha_{0} {=} -\sum_{i=1}^{r}n_{i}\alpha_{i},$ and $ n_0 {=} 1.$
The field $\phi^{a}$ ($a {=} 1,\cdots,r$) is $a$-th component of the scalar
field $\Phi$, and $\alpha_{i}$ ($i {=} 1,\cdots,r$) are simple roots of the
algebra $g$. The $m$ sets the mass scale and the integers $n_i$ are the
so-called Kac labels which are defined for each Lie algebra.

To extract the boundary reflection amplitudes in the framework of
the Feynman's perturbation theory, two-point Green's functions
are considered in the coordinate space rather than in the momentum space.
At one-loop order, there are three types of the relevant Feynman diagrams.

\begin{picture}(350,175)(-50,-110)
\thicklines
\put(0,0){\circle{40}}
\put(-20,50){\line(0,-1){100}}
\put(-15,35){a}
\put(-15,-35){a}
\put(25,0){b}
\put(-18,-70){Type I.}
\put(170,0){\circle{40}}
\put(130,0){\line(1,0){20}}
\put(130,50){\line(0,-1){100}}
\put(135,35){a}
\put(135,-35){a}
\put(137,5){b}
\put(195,0){c}
\put(135,-70){Type II.}
\put(300,0){\circle{40}}
\put(300,20){\line(0,1){30}}
\put(300,-20){\line(0,-1){30}}
\put(305,35){a}
\put(305,-35){a}
\put(270,0){b}
\put(325,0){c}
\put(280,-70){Type III.}
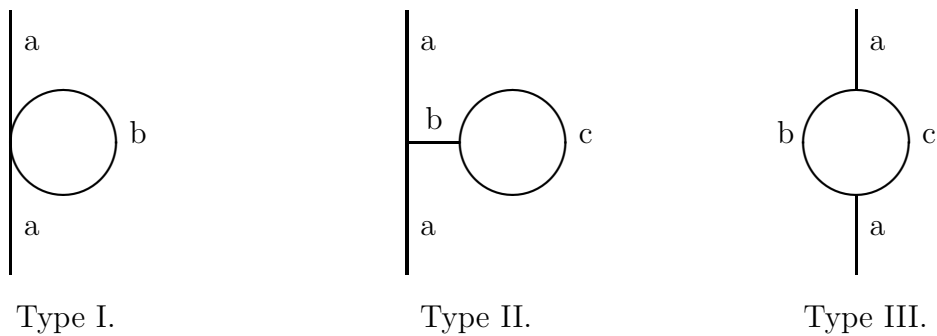
\put(10,-100){Figure 1. Diagrams for the one-loop two-point function.}
\end{picture}

After the infinite as well as finite mass renormalization, the remaining
terms in the two-point Green's function at one-loop order can be written
in the following form in the asymptotic region up to exponentially damped
term as $x, x'$ tend to $ -\infty$ away from the boundary\cite{Kim1}:
\begin{equation}
\int \frac{dw}{2 \pi} e^{-iw(t'-t)} \frac{1}{2 \bar{k}}
 \left(  e^{i \bar{k} |x'-x|} +K_a(w)  e^{-i \bar{k} (x'+x)} \right),
  ~~~~ \bar{k}=\sqrt{w^2-m_a^2}
\end{equation}
Two particle amplitudes of the elastic boundary reflection matrix are defined
as the coefficients $K_a$ of the reflection term and $K_a(\theta)$ is obtained
using $w {=} m_a ch\theta$.
Each contribution to $K_a(\theta)$ from three types of the diagrams
depicted in figure 1 are listed below\cite{Kim1,Kim2}:
\begin{eqnarray}
K_a^{(I)}(\theta) &=& \frac{1}{4 m_a sh\theta} 
 \left(\frac{1}{2 \sqrt{m_a^2 sh^2\theta+m_b^2}} +\frac{1}{2 m_b} \right)
 ~C_1 ~S_1
\label{K-I}  \\
K_a^{(II)}(\theta) &=& \frac{1}{4 m_a sh\theta}
 \left(\frac{-i}{(4 m_a^2 sh^2\theta +m_b^2) 2 \sqrt{m_a^2 sh^2\theta+m_c^2}}
  +\frac{-i}{ 2 m_b^2 m_c} \right) ~C_2 ~S_2
\label{K-II}  \\
\label{K-III}
K_a^{(III)}(\theta) &=& \frac{i}{4 m_a sh\theta} ~ C_3 ~S_3 \\
 & & \left(
   \frac{ cos\theta_{ab}^c }{4 m_a m_b^2 (ch^2\theta-cos^2\theta_{ab}^c) }
 - \frac{ m_a ch^2\theta + m_b cos \theta_{ab}^c }
{2 m_a m_b^2 2\sqrt{m_a^2 sh^2\theta +m_c^2}(ch^2\theta-cos^2\theta_{ab}^c)} 
  \right. \hspace{1cm} \nonumber \\
 & & \left. 
 + \frac{ cos\theta_{ac}^b }{4 m_a m_c^2 (ch^2\theta-cos^2\theta_{ac}^b) }
 - \frac{ m_a ch^2\theta + m_c cos \theta_{ac}^b }
{2 m_a m_c^2 2\sqrt{m_a^2 sh^2\theta +m_b^2}(ch^2\theta-cos^2\theta_{ac}^b)}
     \right) \nonumber
\end{eqnarray}
where $\theta_{ab}^c$ is the usual fusing angle defined by
$ m_c^2 {=} m_a^2+m_b^2 +2 m_a m_b ~cos\theta_{ab}^c $ and
$C_i, S_i$ denote numerical coupling factors and symmetry factors,
respectively.
All the expressions in (\ref{K-I},\ref{K-II},\ref{K-III})
have in general non-meromorphic terms when
a theory under consideration has a mass spectrum with more than one mass.
Non-trivial cancellation of the non-meromorphic terms against one another
was first observed explicitly in \cite{KimCho}.

For $a_{2n}^{(2)}$ theories, the classical masses are
\begin{equation}
m_a = 2 \sqrt{2} m ~sin\frac{a\pi}{h} , ~~~~a=1,\dots,n
\end{equation}
where $h {=} 2n {+} 1$ is the Coxeter number for $a_{2n}^{(2)}$
and the non-vanishing three-point couplings are
\begin{equation}
c_{abc}=\cases{\frac{\beta}{\sqrt{2h}} m_a m_b m_c,~~ & if ~~$ a+b+c=h; $ \cr
       -\frac{\beta}{\sqrt{2h}} m_a m_b m_c,~~  & if ~~$ a \pm b \pm c=0 $ }
\label{TPC}
\end{equation}
The four-point couplings can be obtained via a recursion relation as
follows\cite{CM,BCDS3}:
\begin{equation}
c_{abcd} = \frac{\beta^2}{m^2 h} m_{ab}^2 m_{cd}^2
 + \sum_{f} c_{abf} \frac{1}{m_f^2} c_{\bar{f}cd}
\end{equation}

As a specific case, $a_6^{(2)}$ theory is considered here.
This theory has three particles and their single particle reflection
amplitudes at one-loop order are evaluated as
follows\footnote{Intermediate steps are omitted here.
See for examples \cite{Kim1,ChoKim}}:
\begin{displaymath}
K_1 (\theta) = 1+ \frac{i \beta^2}{4 h} \left(
 \frac{-1 sh\theta}{ch\theta-cos\frac{0}{14}\pi}
 +\frac{-1/2 sh\theta}{ch\theta-cos\frac{5}{14}\pi}
 +\frac{1 sh\theta}{ch\theta-cos\frac{7}{14}\pi}
 +\frac{-1/2 sh\theta}{ch\theta-cos\frac{9}{14}\pi} \right.
\end{displaymath}
\begin{equation}
\label{Pert}
\left. + \frac{1 sh\theta}{ch\theta-cos\frac{12}{14}\pi} \right) + O(\beta^4)
\hspace{3.7cm}
\end{equation}
\begin{displaymath}
K_2 (\theta) = 1+ \frac{i \beta^2}{4 h} \left(
 \frac{-1 sh\theta}{ch\theta-cos\frac{0}{14}\pi}
 +\frac{-1 sh\theta}{ch\theta-cos\frac{2}{14}\pi}
 +\frac{-1/2 sh\theta}{ch\theta-cos\frac{3}{14}\pi}
 +\frac{1 sh\theta}{ch\theta-cos\frac{7}{14}\pi} \right.
\end{displaymath}
\begin{displaymath}
\hspace{2cm} \left.  +\frac{1 sh\theta}{ch\theta-cos\frac{10}{14}\pi}
       +\frac{-1/2 sh\theta}{ch\theta-cos\frac{11}{14}\pi}
       +\frac{1 sh\theta}{ch\theta-cos\frac{12}{14}\pi} \right) + O(\beta^4)
\end{displaymath}
\begin{displaymath}
K_3 (\theta) = 1+ \frac{i \beta^2}{4 h} \left(
 \frac{-1 sh\theta}{ch\theta-cos\frac{0}{14}\pi}
 +\frac{-1/2 sh\theta}{ch\theta-cos\frac{1}{14}\pi}
 +\frac{-1 sh\theta}{ch\theta-cos\frac{2}{14}\pi}
 +\frac{-1 sh\theta}{ch\theta-cos\frac{4}{14}\pi} \right.
\end{displaymath}
\begin{displaymath}
\hspace{2.9cm} \left.  +\frac{1 sh\theta}{ch\theta-cos\frac{7}{14}\pi}
   +\frac{1 sh\theta}{ch\theta-cos\frac{8}{14}\pi}
   +\frac{1 sh\theta}{ch\theta-cos\frac{10}{14}\pi}
   +\frac{1 sh\theta}{ch\theta-cos\frac{12}{14}\pi} \right.
\end{displaymath}
\begin{displaymath}
\left. +\frac{-1/2 sh\theta}{ch\theta-cos\frac{13}{14}\pi} \right) + O(\beta^4)
\hspace{4.0cm}
\end{displaymath}
where $h {=} 7$ for $a_{6}^{(2)}$.
Non-meromorphic terms exactly cancel out against one another.

\sect{3. Exact Boundary Reflection Matrix}
For the present purpose, $a_{2n}^{(2)}$ ATFT can
be best understood in terms of a $Z_2$-reduction of $a_{2n}^{(1)}$ theory,
where $i$-th simple root is identified as $(h {-} i)$-th simple roots while
leaving $\alpha_0$ unchanged. The parent theory consists of $n$-particles
as well as their complex conjugates. Upon the reduction, a half of
the spectrum is discarded and the remaining particles become real scalars.
So the $S$-matrices for $a_{2n}^{(2)}$
theory are crossing-symmetric and is given by\cite{AFZ,FKM,CM}
\begin{equation}
S_{ab}(\theta) = \prod_{|a-b|+1 ~ {\rm step} 2}^{a+b-1}\{p\} \{h {-} p\}
\label{SS}
\end{equation}
where
\begin{equation}
(x)=\frac{ sh(\theta/2 +i\pi x /2h) }{ sh(\theta /2 -i\pi x /2h) }, ~~~
\{x\}=\frac{(x-1)(x+1)}{(x-1+2 B)(x+1-2 B)}
\label{BLOCK}
\end{equation}
and the coupling dependence enters via the universal function:
$B(\beta) {=} \beta^2 / (\beta^2 +4\pi)$.
And non-vanishing three-point couplings shown in (\ref{TPC}) can be
read off from the three-point couplings $c_{abc}$ of $a_{2n}^{(1)}$
theory by mapping any of the particle indices $a$ with $a {>} n$ in $c_{abc}$
to $(h {-} a)$ if it exists. This implies the two models share the same
matrix $J(\theta)$ defined by
\begin{equation}
J_a(\theta)=\sqrt{K_a(\theta)/K_{\bar{a}}(i\pi +\theta)}
=K_a(\theta)/\sqrt{S_{aa}(2 \theta)}
\label{Jdef}
\end{equation}
which is introduced in \cite{KY} so that it satisfies a simple
equation for the boundary bootstrap:
\begin{equation}
J_c(\theta)=
J_a(\theta+i {\bar \theta_{ac}^{b}}) J_b(\theta -i {\bar \theta_{bc}^{a}} )
\label{NBB}
\end{equation}
$\theta_{ab}^{c}$ is the fusing angles and ${\bar \theta} {=} i\pi {-} \theta$.
In terms of $a_{2n}$ root systems of the parent theory, $J_i(\theta)$ for
$a_{2n}^{(2)}$ theory (or $a_{2n}^{(1)}$ theory) is given by\cite{KY}
\begin{equation}
J_{b}(\theta)=\prod_{p=0}^{h-1} \left[ 2p+1/2+\epsilon_{b}
\right]^{1/2 \sum_{a} (\lambda_a \cdot w^{-p} \phi_b)}
\label{JJ}
\end{equation}
where
\begin{equation}
 [ x ] = \frac{ (x-1/2) (x+1/2)} {(x-1/2+B) (x+1/2-B)}
\label{HBlock}
\end{equation}
$\epsilon_b$ is defined as follows depending on the \lq colour' of $\alpha_b$:
$ \epsilon_{\bullet} {=} 1,~\epsilon_{\circ} {=} 0. $
$\lambda_a$ are dual vectors such that
$(\lambda_a {\cdot} \alpha_b) {=} \delta_{ab}$. $w$ is the Coxeter element
and positive roots $\phi_b$ are specially chosen representatives of the Weyl
orbits such that $w \phi_b$ are negative roots.

The single particle boundary reflection amplitude $K_i(\theta)$ can now
be obtained from the defining relation (\ref{Jdef}).
In manipulating the building blocks, the following identities are useful:
\begin{equation}
\{x\}_{2\theta}=[x/2]_{\theta}/[h {-} x/2]_{\theta},~~~[2h {+} x]=[x],
~~~[-x]=1/[x]
\label{Iden}
\end{equation}
For an illustration, here $a_{6}^{(2)}$ theory is considered.
The two-particle scattering amplitudes are
\begin{eqnarray}
\label{Ss}
S_{11}(\theta) & = & \{1 \} \{6 \}    \\ 
S_{22}(\theta) & = & \{1 \} \{6 \} \{3 \} \{ 4 \}    \nonumber \\
S_{33}(\theta) & = & \{1 \} \{6 \} \{3 \} \{ 4 \} \{5 \} \{2 \} \nonumber
\end{eqnarray}
and a little amount of work with the $a_{6}$ root space produces
the single particle amplitudes $J_i(\theta)$:
\begin{eqnarray}
\label{Js}
J_1(\theta) &=&  [\frac{1}{2}] ^{\frac{1}{2}} [\frac{3}{2}] ^{\frac{2}{2}} 
 [\frac{5}{2}] ^{\frac{2}{2}} [\frac{7}{2}] ^{\frac{2}{2}} 
 [\frac{9}{2}] ^{\frac{2}{2}} [\frac{11}{2}] ^{\frac{2}{2}} 
 [\frac{13}{2}] ^{\frac{1}{2}}   \\ 
J_2(\theta) &=&  [\frac{1}{2}] ^{\frac{1}{2}} [\frac{3}{2}] ^{\frac{3}{2}} 
 [\frac{5}{2}] ^{\frac{4}{2}} [\frac{7}{2}] ^{\frac{4}{2}} 
 [\frac{9}{2}] ^{\frac{4}{2}} [\frac{11}{2}] ^{\frac{3}{2}} 
 [\frac{13}{2}] ^{\frac{1}{2}}   \nonumber  \\
J_3(\theta) &=&  [\frac{1}{2}] ^{\frac{1}{2}} [\frac{3}{2}] ^{\frac{3}{2}} 
 [\frac{5}{2}] ^{\frac{5}{2}} [\frac{7}{2}] ^{\frac{6}{2}} 
 [\frac{9}{2}] ^{\frac{5}{2}} [\frac{11}{2}] ^{\frac{3}{2}} 
 [\frac{13}{2}] ^{\frac{1}{2}}  \nonumber
\end{eqnarray}
Inserting $S$ and $J$ of (\ref{Ss},\ref{Js}) into (\ref{Jdef}),
one obtains $K_i(\theta)$, using the identities given in (\ref{Iden}):
\begin{eqnarray}
\label{Exact}
K_1(\theta) &=&  [\frac{1}{2}] [\frac{3}{2}] [\frac{5}{2}] [\frac{7}{2}]
 [\frac{9}{2}] [\frac{11}{2}] \sqrt{ \frac{[3]}{[4]} } \\
K_2(\theta) &=&  [\frac{1}{2}] [\frac{3}{2}]^2 [\frac{5}{2}]^2 [\frac{7}{2}]^2
 [\frac{9}{2}]^2 [\frac{11}{2}] \sqrt{ \frac{[2] [3]}{[5] [4]} } \nonumber \\
K_3(\theta) &=&  [\frac{1}{2}] [\frac{3}{2}]^2 [\frac{5}{2}]^3 [\frac{7}{2}]^3
 [\frac{9}{2}]^2 [\frac{11}{2}] \sqrt{ \frac{[1] [2] [3]}{[6] [5] [4]} }
 \nonumber
\end{eqnarray}
It is easy to see a complete agreement between
the exact result given in (\ref{Exact}) and the perturbative result
given in (\ref{Pert}) by using the following identity:
\begin{equation}
\frac{(x)}{(x \pm B)} =
 1 \mp \frac{i \beta^2}{4 h} \frac{sh\theta}{ch\theta-cos\frac{x}{h} \pi}
 +O(\beta^4)
\label{Comp}
\end{equation}

\sect{4. Conclusions}
In this paper, for $a_{2n}^{(2)}$ affine Toda field theory defined
on a half line, single particle amplitudes of the exact boundary reflection
matrix corresponding to the Neumann boundary condition 
are constructed, hinted by the reduction idea combined with the recently
introduced matrix $J(\theta)$. Specifically, for $a_{6}^{(2)}$
theory, the reflection amplitudes are evaluated in the formulation
developed in \cite{Kim1} and then a hypothesised expression
of $J(\theta)$ is obtained and tested against the one-loop result.
One of the distinguished features of the reflection amplitudes corresponding
to the Neumann boundary condition for $a_{2n}^{(2)}$ theory is
the fact that they exhibit a new kind of square root singularity.

$a_{2n}^{(2)}$ affine Toda field theory is unique in the sense that
despite being based on twisted algebras, their mass ratios do remain fixed
under quantum corrections. This is deeply related to the fact
that their root lattices are self-dual.
If the square root singularities were simple poles, they would signal
the existence of boundary states\cite{GZ} because they are inexplicable
in terms of bulk three-particle vertices, but it is believed that no boundary
states are allowed to exist with the Neumann boundary condition.
In practice, the existence of the square root singularity for
$a_{2n}^{(2)}$ theory slows down
the rate of change of the phases of single particle reflection amplitudes
in the physical region (positive real line) of the complex rapidity plane,
compared with those of the parent theory.

At this stage, it is not clear whether the new type of the square root
singularities would appear elsewhere.
It would be very interesting to find other solutions
of the boundary Yang-Baxter equation, the boundary bootstrap equation
and the boundary crossing-unitarity relation,
which exhibit the new type of the square root singularities.

It is remarked a few striking features of integrable
relativistic quantum field theory defined on a half line.
Firstly, the single particle amplitudes of boundary reflection matrix
at one-loop order have no ambiguity of divergences arising from such as
tadpoles at all.
Secondly, given a way to compute reflection amplitudes directly
without the use of $S$-matrices as in the formulation
developed in \cite{Kim1}, one can derive scattering amplitudes
via the boundary crossing-unitarity relation, not calculating them directly.
In this case, $n$-loop single particle amplitudes of
boundary reflection matrix correspond to $(n {-} 1)$-loop two-particle
amplitudes of scattering matrix.
Thirdly, boundary reflection amplitudes can have, due to three-particle
vertices in the bulk, a rich structure of poles which should not be
interpreted as a signal of the existence of boundary
states\footnote{A formal possibility of this picture has been discussed
in \cite{CDRS}.}.

\section*{Acknowledgement}
One(JDK) of the authors wishes to thank Dr. H.S. Cho for useful discussions.


\newpage
\begin{thebibliography}{99}
\bibitem{ELOP}
R.J. Eden, P.V. Landshoff, D.I. Olive and J.C. Polkinghorne,
The analytic $S$-matrix, (Cambridge University Press 1966).
\bibitem{ZZ}
A.B. Zamolodchikov and Al.B. Zamolodchikov, Ann. Phys. 120 (1979) 253.
\bibitem{BCDS}
H.W. Braden, E. Corrigan, P.E. Dorey and R. Sasaki,
Nucl. Phys. B 356 (1991) 469.
\bibitem{DGZ}
G.W. Delius, M.T. Grisaru and D. Zanon, Nucl. Phys. B 382 (1992) 365.
\bibitem{BS}
H.W. Braden and R. Sasaki, Nucl. Phys. B 379 (1992) 377.
\bibitem{BCKKS}
H.W. Braden, H.S. Cho, J.D. Kim, I.G. Koh and R. Sasaki,
Prog. Theor. Phys. 88 (1992) 1205.
\bibitem{FK}
A. Fring and R. K\"oberle,
Nucl. Phys. B 421 (1994) 159; Nucl. Phys. B 419 (1994) 647.
\bibitem{GZ}
S. Ghoshal and A.B. Zamolodchikov, Int. J. Mod. Phys. A 9 (1994) 3841;
Int. J. Mod. Phys. A 9 (1994) 4353.
\bibitem{Che}
I.V. Cherednik, Theor. Math. Phys. 61 (1984) 977.
\bibitem{Sas}
R. Sasaki, In the proceedings of the conference
\lq\lq Interface between physics and mathematics",
Hangzhou, China, 6-17 September 1993 (World Scientific 1994).
\bibitem{AK}
C. Ahn and W.M. Koo, \lq\lq Exact boundary $S$-matrices of the supersymmetric
sine-Gordon theory on a half line", EWHA-TH-006, hep-th/9509056.
\bibitem{SSW}
H. Saleur, S. Skorik and N.P. Warner, Nucl. Phys. B441 (1995) 421.
\bibitem{AKL}
M. Ameduri, R. Konik and A. LeClair, Phys. Lett. B354 (1995) 376.
\bibitem{CDRS} E. Corrigan, P.E. Dorey, R.H. Rietdijk and R. Sasaki,
Phys. Lett. B 333 (1994) 83.
\bibitem{CDR} E. Corrigan, P.E. Dorey and R.H. Rietdijk,
Supplement of Prog. Theor. Phys. 118 (1995) 143.
\bibitem{Mac}
A. MacIntyre, J. Phys. A 28 (1995) 1089.
\bibitem{BCDR}
P. Bowcock, E. Corrigan, P.E. Dorey and R.H. Rietdijk,
Nucl. Phys. B 445 (1995) 469.
\bibitem{PZ}
S. Penati and D. Zanon, Phys. Lett. B 358 (1995) 63.
\bibitem{MS}
M.F. Mourad and R. Sasaki, \lq\lq Nonlinear sigma models on a half plane",
hep-th/9509153.
\bibitem{IOZ}
T. Inami, S. Odake and Y.Z Zhang, Phys. Lett. B 359 (1995) 118.
\bibitem{PRZ}
S. Penati, A. Refoli and D. Zanon, IFUM-518-FT, hep-th/9510084; IFUM-522-FT,
hep-th/9512174.
\bibitem{Kim1}
J.D. Kim, Phys. Lett. B 353 (1995) 213.
\bibitem{Kim2}
J.D. Kim, \lq\lq Boundary reflection matrix for $ade$ affine Toda field
theory", DTP-95/31, hep-th/9506031, to appear in J. Phys. A (1996).
\bibitem{Kim3}
J.D. Kim, Phys. Rev. D 53 (1996) 4441.
\bibitem{KY}
J.D. Kim and Y. Yoon, \lq\lq Root systems and boundary bootstrap",
KAIST/THP-96/701, hep-th/9603111.
\bibitem{OT}
D.I. Olive and N. Turok, Nucl. Phys. B 215 (1983) 470.
\bibitem{CM}
P. Christe and G. Mussardo, Int. J. Mod. Phys. A 5 (1990) 4581.
\bibitem{Sas2}
R. Sasaki, Nucl. Phys. B 383 (1992) 291.
\bibitem{BCDS2}
H.W. Braden, E. Corrigan, P.E. Dorey and R. Sasaki, Nucl. Phys. B 338
(1990) 689.
\bibitem{KimCho}
J.D. Kim and H.S. Cho, \lq\lq Boundary reflection matrix for $D_4^{(1)}$
affine Toda field theory", DTP/95-23, hep-th/9505138.
\bibitem{BCDS3}
H.W. Braden, E. Corrigan, P.E. Dorey and R. Sasaki, Aspects of affine
Toda field theory, in Proc. 10th Winter School on Geometry and Physics,
Srni, Czechoslovakia; Integrable Systems and Quantum Groups, Pavia, Italy;
Spring Workshop on Quantum Groups, ANL, USA.
\bibitem{ChoKim}
H.S. Cho and J.D. Kim, \lq\lq Perturbative verification of the boundary
reflection matrix for $a_n^{(1)}$ affine Toda field theory", KAIST/THP-95/702.
\bibitem{AFZ}
A.E. Arinshtein, V.A. Fateev and A.B. Zamolodchikov,
Phys. Lett. B 87 (1979) 389.
\bibitem{FKM}
P.G.O. Freund, T. Klassen and E. Melzer, Phys. Lett. B 229 (1989) 243.
\end{thebibliography}
\end{document}